# Probability Distribution of a Passive Scalar in Isotropic Turbulence

Zheng Ran, Xingjie Yuan and Yaoyao Wang
Shanghai Institute of Applied Mathematics and Mechanics,
Shanghai University, Shanghai 200072, P.R.China

In this letter, we present developments of the Hamiltonian approach to problems of the probability distribution for a passive scalar in isotropic turbulence, and also considers specific applications of the modified Prelle-Singer procedure to turbulence models. The following key questions are discussed and solved: what is the general dynamical structure of the resulting scale equation permitted by passive scalar turbulence models? What are the general requirements of the relations between canonical variables and the canonical variabes representation for turbulence by using canonical variables? It is shown that the existence of the Haniltonian representation in turbulence is a privilege of only turbulence systems for which the variational principle of least action is impossible The master equation of the probability distribution of a passive scalar in isotropic turbulence can also be deduced explicitly.

There is presently no fully deductive theory which starts from the Navier-Stokes equation and leads to the some basic experimental laws [1]. It has always been our opinion that a statistical theory of turbulence required for a sound foundation, based directly on the equations of motion, just as the statistical mechanics of dynamical systems is based on Hamilton-Jacobi. It is logical to apply the techniques of classical statistical mechanics to the turbulence problem of fluid dynamics. Considerable progress has been made in this respect during the last years but there are still many important problems to be solved before a successful statistical theory of turbulence can be developed [2]. In particular, there is no systematic derivation of the probability distribution of turbulence fields [3]. We outline here a new Hamiltonian approach to problems of the passive scalar in istropic turbulence. A recent paper by Pradeep et al [4] helped to motivate our work and point to a fruitful line of attack. In a nice paper, Gladwin Pradeep et al present the details of the mathematical backgrouds. By using the modified Prelle-Singer procedure [5], they consider a second order nonlinear ordinary equation, and they identify five new integrable cases. Among these five, four equations admit time-dependent first integrals and remaining one admits time-independent first integral. From the time-independent first integral, nonstandard Hamiltonian stucture is deduces, thereby proving the Liouville sense of integrability. Based on this mathematical approach, a Gibbsian statistical mechanics of turbulence may be developed by taking the scale equation as the dynamical system analogous to the canonical variables of Hamilonian systems.

We shall start from the equation for the diffusion and convection of the scalar quantity $\Gamma$ per unit of mass [6]

$$\frac{\partial \Gamma}{\partial t} + U_i \frac{\partial \Gamma}{\partial x_i} = \frac{\partial}{\partial x_i}\left(\bar{f} \frac{\partial \Gamma}{\partial x_i}\right) + F_\gamma, \tag{1}$$

where $\bar{f}$ denotes the molecular-transport coefficient. $F_\gamma$ is a driving force or source. If we assume that $\bar{f}$ is constant and if we put

$$\Gamma = \bar{\Gamma} + \gamma, \tag{2}$$

$$U_i = \bar{U}_i + u_i, \tag{3}$$

Keeping in mind that $\bar{\Gamma}$ and $\bar{U}_1$ are constant throughout the field, then we have

$$\frac{\partial \gamma}{\partial t} + \bar{U}_1 \frac{\partial \gamma}{\partial x_1} + u_i \frac{\partial \gamma}{\partial x_i} = \bar{f} \frac{\partial^2 \gamma}{\partial x_l \partial x_l}. \tag{4}$$

As in the case of turbulent velocity field, the correlations between varying quantities at two points of the field are very useful for describing geometrical relations in the structure of the field. Let us first consider the double correlation

$$[Q_{\gamma,\gamma}(r,t)]_{A,B} = \overline{\gamma_A \gamma_B}, \tag{5}$$

with its coefficient

$$[R_{\gamma,\gamma}(r,t)]_{A,B} = \frac{\overline{\gamma_A \gamma_B}}{\gamma'^2} \equiv f(r,t), \tag{6}$$

where $\gamma_A$ and $\gamma_B$ are the values of $\gamma$ at points A and B, respectively, located at a distance $r$ from each other, and $\gamma' = \sqrt{\overline{\gamma_A^2}} = \sqrt{\overline{\gamma_B^2}}$ is the root-mean-square value or the intensity of the $\gamma$-fluctuations. Apart from the double correlation between $\gamma$ at two points, we also have correlations between $\gamma$ at one point and a velocity component at another point. $k_\gamma(r,t)$ is the coefficient of the spatial correlation between $\gamma_B, \gamma_A$ and the velocity component $(u_r)_A$ at A in the durection of $r$:

$$k_\gamma(r,t) = \frac{\overline{\gamma_B \gamma_A (u_r)_A}}{\gamma'^2 u'} \equiv h(r,t), \tag{7}$$

The dynamical equation for scalar turbulence may be written in terms of the correlation coefficients [6],

$$\frac{\partial}{\partial t}(\theta f) - 2\theta u'\left(\frac{\partial h}{\partial r} + \frac{4h}{r}\right) = 2\bar{f}\theta\left(\frac{\partial^2 f}{\partial r^2} + \frac{4}{r}\frac{\partial f}{\partial r}\right) \tag{8}$$

where $(r,t)$ is the spatial and time coordinates, $u'$ denotes the intensity of the turbulent fluctuation. $\nu$ is the kinematic viscosity, and $\theta = \overline{\gamma'^2}$ denotes the turbulence intensity, $\bar{f}$ is the molecular transport coefficient. We may apply to the dynamic equation for $f(r,t)$, the same procedure for studyng a possible behaviour of self-preservation of the scalar field as we did for the velocity field [7]. Following von Karman and Howarth [6], we introduce the new variables

$$\xi = \frac{r}{l(t)} \tag{9}$$

where $l = l(t)$ is a uniquely specified similarity length scale.

We may apply to the dynamic equation for $f(r,t), h(r,t)$, the same procedure for studying a possible behaviour of self-preservation of the scalar field as we did for the velocity field [9]. We obtain from Eq. (8):

$$12f + \frac{\lambda_\gamma^2}{\bar{f}} \cdot \frac{1}{l} \cdot \frac{dl}{dt} \cdot \xi \frac{df}{d\xi} + 2\frac{u'\lambda_\gamma^2}{\bar{f}l} \cdot \left[\frac{dh}{d\xi} + \frac{2}{\xi}h\right] + 2\frac{\lambda_\gamma^2}{l^2} \cdot \left[\frac{d^2f}{d\xi^2} + \frac{2}{\xi}\frac{df}{d\xi}\right] = 0, \tag{10}$$

It is convenient to divide by $2\dfrac{u'\lambda_\gamma^2}{\bar{f}l}$ so that the transformed equation reduces to

$$\frac{6\bar{f}l}{u'\lambda_\gamma^2} \cdot f + \frac{1}{2u'} \cdot \frac{dl}{dt} \cdot \xi \frac{df}{d\xi} + [1]\cdot\left[\frac{dh}{d\xi} + \frac{2}{\xi}h\right] + \frac{\bar{f}}{u'l} \cdot \left[\frac{d^2f}{d\xi^2} + \frac{2}{\xi}\frac{df}{d\xi}\right] = 0, \tag{11}$$

Differentiating (11) with respect to time for constant $\xi$, we obtain

$$\frac{d}{dt}\left[\frac{6\bar{f}l}{u'\lambda_\gamma^2}\right]\cdot f + \frac{d}{dt}\left[\frac{1}{2u'}\cdot\frac{dl}{dt}\right]\cdot\xi\frac{df}{d\xi} + \frac{d}{dt}\left[\frac{\bar{f}}{u'l}\right]\cdot\left[\frac{d^2f}{d\xi^2} + \frac{2}{\xi}\frac{df}{d\xi}\right] = 0. \tag{12}$$

We shall analyse the following possible case associated with above equation. There exists only one independent linear relation with constant coefficients of the type

$$c_1 \cdot f + c_2 \cdot \xi\frac{df}{d\xi} + c_3 \cdot \left[\frac{d^2f}{d\xi^2} + \frac{2}{\xi}\frac{df}{d\xi}\right] = 0, \tag{13}$$

in which not all $c_1, c_2,$ and $c_3$ equal zero.

Since $c_3 \neq 0$, the equation can be written in the form

$$\frac{d^2f}{d\xi^2} + \left(\frac{4}{\xi} + \frac{a_1}{2}\xi\right)\frac{df}{d\xi} + \frac{a_2}{2}f = 0, \tag{14}$$

with boundary conditions $f(0) = 1$, $f(\infty) = 0$, and $a_1$ and $a_2$ are constant coefficients.

We find from (12) and (14) that

$$\frac{d}{dt}\left[\frac{1}{2u'}\cdot\frac{dl}{dt}\right]-\frac{a_1}{2}\cdot\frac{d}{dt}\left[\frac{\bar{f}}{u'l}\right]=0, \tag{15}$$

$$\frac{d}{dt}\left[\frac{6\bar{f}l}{u'\lambda_\gamma^2}\right]-\frac{a_2}{2}\cdot\frac{d}{dt}\left[\frac{\bar{f}}{u'l}\right]=0. \tag{16}$$

In fact, the problem is thus reduce to solving a soluable system, furthermore, let $z=\frac{1}{l^2}$, one can prove that there is a self-closed second order nonlinear dynamical system for $z=z(t)$:

$$\frac{d^2z}{dt^2}-\left(\frac{3}{2}\right)\cdot z^{-1}\cdot\left[\frac{dz}{dt}\right]^2+\left(a_1\bar{f}+\frac{a_2}{2}v\right)\cdot z\cdot\left[\frac{dz}{dt}\right]+(a_1a_2\bar{f}v)\cdot z^3=0. \tag{17}$$

The most important finds is that :

By using the modified Prelle-Singer procedure [4,5], we indentify the new integrable cases in this equation. Among these cases, four equations admit time dependent first integrals and the remaining one admits time-independent first integral. From the time-independent first integral, nonstandard Hamiltonian structure is deduced, thereby proving the Liouville sense of integrability. In the case of time-dependent integrals, we either explicitly integrate the system or transform to a time-independent case and deduce the underlying Hamitonian structure. We also demonstrate that the above second order ordinary differential equation is intimately related t the Liouville equation. To further consideration the following variable is introduced:
where

$$k_1=-\frac{3}{2}, \tag{18}$$

$$k_3=a_1\bar{f}+\frac{a_2}{2}v, \tag{19}$$

$$k_4=a_1a_2\bar{f}v. \tag{20}$$

By using the modified Prelle-Singer procedure, we indentify the new integrable cases in th is equation. From the time independent first integral, nonstatandard Hamiltionian strucuture is deduced thereby proving the Liouville sense of integrability.
The paprameter choice condition read as

$$\Delta\equiv k_3^2-4k_4(2+k_1)$$
$$=\left[\frac{1}{2}a_2v-a_1\bar{f}\right]^2 \tag{21}$$
$$>0$$

We use the canonical transformation

$$z=\frac{U}{P}, \tag{22}$$

$$p=\frac{1}{2}P^2. \tag{23}$$

where $p$ is the canonical momenta defined by

$$p = z^{(2-r)k_1} \cdot \left\{ \frac{dz}{dt} + \frac{k_3(r-1)z^2}{r(2+k_1)} \right\}^{1-r}. \tag{24}$$

where

$$r = \frac{k_3^2 \pm k_3\sqrt{k_3^2 - 4k_4(2+k_1)}}{2k_4(2+k_1)}, \tag{25}$$

For the choice $\Delta > 0$, the Hamiltonian can be written in terms of the new cnaonical variables as

$$H = 2(2+k_1)rr_{12}\left[P^{2+k_1}U^{-k_1}\right]^{r_{12}} - 2^{r_{12}}k_3(r-1)U^2. \tag{26}$$

The associated canonical equations of motion now become

$$\frac{dU}{dt} = \frac{2(2+k_1)^2 rr_{12}^2}{P}\left[\frac{P^{2+k_1}}{U^{k_1}}\right]^{r_{12}}, \tag{27-a}$$

$$\frac{dP}{dt} = 2^{r_{12}}k_3(r-1)2U + \frac{2k_1(2+k_1)rr_{12}^2}{U}\left[\frac{P^{2+k_1}}{U^{k_1}}\right]^{r_{12}}. \tag{27-b}$$

Where

$$r_{12} = \frac{r-1}{r-2}. \tag{28}$$

Rewritng (27-a) for

$$P = \left[\frac{U^{k_1 r_{12}}U'}{2(2+k_1)^2 rr_{12}^2}\right]^{\frac{1}{(2+k_1)r_{12}-1}}, \tag{29}$$

And substituting the latter into (26) we get

$$H = \mu_2 U^{m_1}[U']^{m_2} + \mu_3 U^2 \equiv E. \tag{30}$$

where

$$\mu_2 = \frac{2(2+k_1)rr_{12}}{\left[2(2+k_1)^2 rr_{12}^2\right]^{m_3(2+k_1)r_{12}}}, \tag{31}$$

$$\mu_3 = -2^{r_{12}}k_3(r-1), \tag{32}$$

$$m_1 = k_1[(2+k_1)m_3 - r_{12}], \tag{33}$$

$$m_2 = r_{12}(2+k_1)m_3. \tag{34}$$

Which in turn can be brought to the form

$$\frac{dU}{dt} = \left\{\frac{E - \mu_3 U^2}{\mu_2 U^{m_1}}\right\}^{\frac{1}{m_1}}. \tag{35}$$

Now integrating the above equation we get

$$t - t_0 = \frac{m_2 U}{m_1 + m_2} \left[ \frac{\mu_2 U^{m_1}}{E} \right]^{\frac{1}{m_2}} F\left[ \frac{m_1 + m_2}{2m_2}, \frac{1}{m_2}, \frac{m_1 + 3m_2}{2m_2}, \frac{\mu_3 U^2}{E} \right]. \qquad (36)$$

where $F(a,b,c,x)$ is the hypergeometric function and $t_0$ is an integration constant.

The above investigation will help us to reveal the turbulence problem from the standpoint of classical statistical mechanics. It is well known that : The application of Gibbs statistical mechanics to turbulence has been discussed by Burgers, Onsager, Hopf, and other authors [2]. A Gibbsian statistical mechanics of turbulence may be developed by taking the real and imaginary parts of the wave vector components of the velocity field as phase space coordinates analogous to the canonical variables of Hamiltonian systems. Here, our work directly in terms of the canonical variables, which are written $(U, P)$ in general sense. This is the phase space. In the Gibbs statisitial mechanics of conservative systems satisfying Liouville's theorm. These conservation laws are sfficient to construct the probability distribution. Conservation of probability in phase space $F = F(t, U, P)$ requires that satisfy

$$\frac{dF}{dt} = \frac{\partial F}{\partial t} + \frac{dU}{dt} \cdot \frac{\partial F}{\partial U} + \frac{dP}{dt} \cdot \frac{\partial F}{\partial P} = 0. \qquad (37)$$

This is called the Liouville-like equation for scalar turbulence. Combining this equation with the Hamiltonian canonical equations (27), we could get the general time-dependent Liouville equation as the master equation for the scalar turbulence, which provide the probability description of the scalar turbulence.

## ACKNOWLEDGEMENTS

The work was supported by the National Natural Science Foundation of China (Grant Nos.11172162, 10572083).